\pdfoutput=1

\documentclass[aps,prl,a4paper,reprint,superscriptaddress,showpacs]{revtex4-1}
\usepackage{graphicx,graphics,color,epsfig}
\usepackage{hyperref}
\usepackage{amsmath}
\usepackage{amsfonts}
\usepackage{amssymb}
\usepackage{appendix}

\begin{document}
\title{Generation of a fully valley-polarized current in bulk graphene}
\author{Yu Song}
\email{kwungyusung@gmail.com}
\affiliation{Institute of Electronic Engineering, China Academy of Engineering Physics,
Mianyang 621900, P. R. China}
\author{Feng Zhai}
\email{fzhai@zjnu.cn}
\affiliation{Center for Statistical and Theoretical Condensed Matter Physics and
Department of Physics, Zhejiang Normal University, Jinhua 321004, P. R. China}
\author{Yong Guo}
\affiliation{Department of Physics and State Key Laboratory of Low-Dimensional Quantum
Physics, Tsinghua University, Beijing 100084, P. R. China}

\begin{abstract}
The generation of a fully valley-polarized current (FVPC) in bulk
graphene is a fundamental goal in valleytronics.
To this end, we investigate valley-dependent transport through a
strained graphene modulated
by a finite magnetic superlattice. It is found that this device
allows a coexistence of insulating transmission gap of one valley
and metallic resonant band of the other. Accordingly, a
substantial bulk FVPC appears in a wide range of edge orientation
and temperature, which can be effectively tuned by structural
parameters. A valley-resolved Hall configuration is designed to
measure the valley polarization degree of the filtered current.
\end{abstract}

\date{\today}
\maketitle

The low-energy electronic elementary excitations in bulk graphene
originate from the out-plane $p_{z}$-$p_{z}$ hybridization.\cite{wallace1947band}
Their massless energy dispersion
is well described by Dirac cones at the six corners of the
Brillouin zone.\cite{semenoff1984condensed} The six cones can be
divided into two inequivalent groups labeled by the valley index
$K$ and $K^{\prime }$. Intervalley coupling or scattering requires
a rather large change of the momentum and is thus suppressed in
clean graphene samples. This
independence suggests that the valley degree of freedom could be
utilized as an information
carrier.\cite{valve,detect,gap,warping1,strain,linedefect}

How to generate a high-contrast
valley population of charge carriers is a fundamental goal of
graphene valleytronics. Several proposed valley filters require
either a point contact with zigzag edges\cite{valve} or a breaking
of the inversion symmetry.\cite{detect,gap} These factors may
break the specific \emph{bulk} elementary excitation that is
essential for most of the excitement about
graphene,\cite{RMP,RMP1} such as
Klein tunneling, nonzero minimum conductivities, and half-integer
quantum Hall effect. Therefore, several schemes of valley
filtering have been proposed based on bulk graphene,
utilizing either valley-dependent trigonal band warping,\cite{warping1}
or pseudo magnetic fields
induced by strain.\cite{strain} However,
the generated valley polarization is shown to be low
even at zero temperature.

\begin{figure}[tb]
\includegraphics[width=0.9\linewidth]{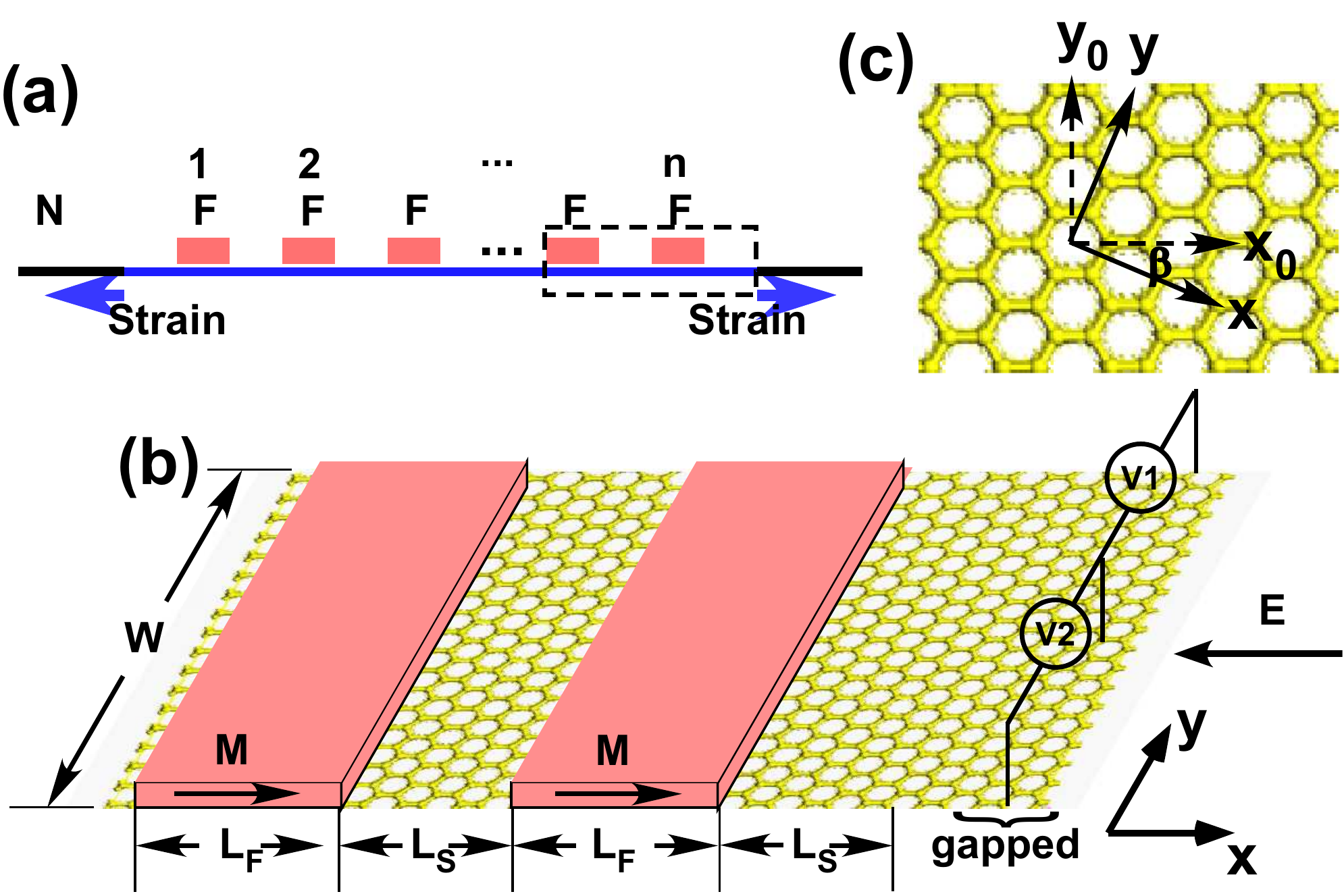}
\caption{ (a) Schematic diagram of the strained graphene modulated
by a finite magnetic superlattice. The bulk graphene sample is placed on a substrate in the ($x$,%
$y$) plane. In the filtering region the substrate is subject to a uniaxial
tension along the $x$-direction. The periodic magnetic field is
created by a finite superlattice of $n$ ferromagnetic metal (FM)
strips depositing on top of the filtering region. (b) The last two
unit cells of the superlattice (top view). To measure the degree
of the valley polarization, a gapped region is connected to the
considered structure from the right. (c) The edge orientation
angle $\protect\beta$. } \label{fig1}\centering
\end{figure}

In this Letter we present a scheme
to achieve a fully valley-polarized current (FVPC) in \emph{bulk}
graphene. The proposed valley filter
is a strained graphene under a
periodic magnetic modulation [see Fig.~\ref{fig1}(a)].
We show that, the combination
between the periodic magnetic field and the strain can lead to a
coexistence of insulating transmission gap of one valley and
metallic resonant band of the other.
Under this mechanism, the bulk
FVPC survives in a wide range of edge orientation, temperature,
and structural parameters. We also discuss how to measure the
valley polarization degree of the output current.

Suppose that the magnetic
superlattice consists of $n$ ferromagnetic metal (FM) strips. Each
FM strip $j$ ($1\leq j\leq n$) has a size $L_{F}$ along the $x$
direction, a magnetization $\mathbf{M}=M\mathbf{e}_{x}$, a
distance $L_{S}$ to the nearest strip(s) [see Fig.~\ref{fig1}(b)],
and is close enough to the graphene plane. The induced magnetic
field can be approximately described\cite{strain} by the vector
potential $\mathbf{A}_{M}(\mathbf{r})=\sum_{j}A_{M}\Theta
( x-x_j^M)\Theta (x_j^S-x)\mathbf{e}_{y}$. Here $\Theta (x)$ is
the Heaviside step function, $x_j^S=jL$, $L=L_{S}+L_{F}$ the size
of a unit cell, and $x_j^M=x_j^S-L_{F}$. The uniaxial tensile
strain is homogeneous in the whole filtering region,
i.e., $u_{yy}-u_{xx}=u$ and $u_{xy}=0$.
It leads to changes in the nearest-neighbor hopping amplitudes, and can be described by
pseudo magnetic vector potentials.\cite{RMP1,pseudo} When the edge
orientation (the $x$-axis) has an angle $\beta $ with respect to
the armchair direction $Ox_{0}$ [see Fig.~\ref{fig1}(c)], the
pseudo vector potential reads $\mathbf{A}_{S}(\mathbf{r})=\mathbf{A}%
_{S}\Theta (x)\Theta (L^{t}-x)$ with $A_{Sy}+iA_{Sx}=A_Se^{i 3\beta}$.
Here $A_S\propto u$ and
$L^{t}=nL+L_{S}$ is the total length of the filtering region.

In the low-energy continuum approximation,
the Hamiltonian for a given valley is\cite{RMP} $ H_{\xi }=v_{F}%
\boldsymbol{\sigma}\cdot (\mathbf{p}+e\mathbf{A}_{M}+\xi \mathbf{A}%
_{S}/v_{F}), $ where $\xi =\pm $ for the valley $K$ and $K^{\prime
}$, $v_{F}$ is the Fermi velocity, $\boldsymbol{\sigma}=(\sigma
_{x},\sigma _{y})$ is the pseudospin Pauli matrices, and
$\mathbf{p}$ is the momentum operator. For brevity, hereafter we
express all quantities in dimensionless form by means of a
characteristic length $l_0=10$ nm and energy unit $E_0=\hbar
v_F/l_0=56.55$ meV. We assume that the sample width $W>>L^t $ so
that edge details are not important.\cite{WL}
For an electron with energy $E$ and incident angle $\alpha$,
the envelope function in each region
$i$ ($i=N,S,M$) of the building block $j$ has the form
\begin{eqnarray}
\psi_{ij}(\mathbf{r})&=&e^{iqy-i\xi
\int_{0}^{x}A_{Sx}(x)dx}[\phi^{i}_{+}(x)u_{ij}+\phi^{i}_{-}(x)v_{ij}],\notag \\
\phi^{i}_{\pm}(x)&=&e^{{\pm}ik_ix}[1,\:(\pm k_{i}+iq_{i})/E_{i}
]^T/\sqrt{2}\label{phi}.
\end{eqnarray}
Here $q=E\sin \alpha$ is the conserved transverse momentum,
$E_{N}=E_{S}=E$, $E_{M}=E-U$ with $U$ being a common gate voltage
on all FM strips, $q_{N}=q$, $q_{S}=q+\xi A_{Sy}$, $
q_{M}=q_{S}+A_{M}$, and $k_{i}=\textmd{sgn}
(E_i)\sqrt{E_{i}^{2}-q_{i}^{2}}$.

The wave amplitudes $u_{ij}$ and
$v_{ij}$ are determined from the continuity of the envelope function
and the scattering boundary condition $u_{N,0}=1$ and
$v_{N,n+1}=0$. We write
\begin{equation}
\left(
\begin{array}{c}
 u_{S,j+1} \\
 v_{S,j+1}
\end{array}\right)=M_{j}\left(
\begin{array}{c}
u_{Sj} \\
v_{Sj}
\end{array}%
\right), M_1=\left(
\begin{array}{cc}
a & b \\
c & d
\end{array}%
\right),
\end{equation}
where $M_j$ ($1\leq j\leq n$) is the transfer matrix for the
superlattice unit cell $j$. By means of the matrix
$U_i(x)=[\phi^{i}_{+}(x),\phi^{i}_{-}(x)]$ and the condition
$\psi_{S,j+1}(x^S_{j})=\psi_{Mj}(x^S_{j})$ and
$\psi_{Mj}(x^M_{j})=\psi_{Sj}(x^M_{j})$, one can express $M_j$
as\cite{tmm}
\begin{equation}
M_j=U_{S}^{-1}(x^S_{j})U_{M}(x^S_{j})U_{M}^{-1}(x^M_{j})U_{S}(x^M_{j}).\label{Mj}
\end{equation}
From Eq.~(\ref{Mj}) we get $detM_j=1$ and the matrix element
$M_j^{11}=a$, $M_j^{12}=\tau^{j-1}b$ with $\tau =e^{-2ik_{S}L}$,
$M_j^{21}=\tau^{1-j}c$, and $M_j^{22}=d$. Here we have used the
identity $U_i(x+l)=U_i(x)diag\{e^{ik_il},e^{-ik_il}\}$. The
transfer matrix for the finite superlattice is
$N_n=M_n...M_{2}M_1=[(a_n, c_n)^T, (b_n, d_n)^T]$. From the
recurrence relation $N_{j+1}=M_{j+1}N_j$ and $N_1=M_1$, we obtain
the matrix elements of $N_n$ \cite{previous}
\begin{equation}
\left\{
\begin{split}
& a_{n}=\frac{1}{2}(F^{n}+G^{n})+\frac{a-\tau d}{2D}(F^{n}-G^{n}), \\
& \frac{b_{n}}{b}=\frac{\tau ^{n-1}c_{n}}{c}=\frac{1}{D}(F^{n}-G^{n}), \\
& \tau ^{n}d_{n}=\frac{1}{2}(F^{n}+G^{n})-\frac{a-\tau d}{2D}(F^{n}-G^{n}).
\end{split}%
\right.   \label{eq1}
\end{equation}%
Here $B=a+\tau d$, $D=\sqrt{B^{2}-4\tau }$, $F=(B+D)/2$, and
$G=(B-D)/2$.

The total transfer matrix
$M=U_N^{-1}(L^t)U_S(L^t)N_nU_S^{-1}(0)U_N(0)$ determines the
valley-resolved transmission coefficient
$t_{\xi}=u_{N,n+1}=1/M^{22}$. The transmission probabilities
read
\begin{equation}
T_{\xi}(E,q)=|t_{\xi}|^2=\left\vert
\frac{4k_{N}k_{S}}{\lambda _{a}a_{n}+\lambda _{b}b_{n}+\lambda
_{c}c_{n}+\lambda _{d}d_{n}}\right\vert ^{2},  \label{eq2}
\end{equation}%
where $\lambda _{a}=[-(k_{N}-k_{S})^{2}-\Delta _{q}^{2}]e^{2ik_{S}L^{t}}$, $%
\lambda _{b}=[k_{N}^{2}-(k_{S}-i\Delta _{q})^{2}]e^{2ik_{S}nL}$, $\lambda
_{c}=[-k_{N}^{2}+(k_{S}+i\Delta _{q})^{2}]e^{2ik_{S}L_{S}}$, $\lambda
_{d}=(k_{N}+k_{S})^{2}+\Delta _{q}^{2}$, and $\Delta _{q}=q_{N}-q_{S}$.

Recent mobility measurements on graphene indicate that the
electron-phonon scattering can be ignored in the temperature range of 10$K$%
-100$K$.\cite{phonon} In this range, the ballistic
valley-resolved conductance is given by the Landau-B\"{u}ttiker
formula\cite{LB}
\begin{equation}
G_{\xi }(E_{F},T_{P})=G_{0}\int dE\frac{-df}{dE}\int_{-|E_{F}|}^{|E_{F}|}T_{%
\xi }(E, q)\frac{dq}{2\pi /W},  \label{eq3}
\end{equation}%
where $f(E)=[1+e^{(E-E_{F})/T_{P}}]^{-1}$ is the Fermi-Dirac distribution
function at the temperature $T_{P}$ and the Fermi energy $E_{F}$, and $%
G_{0}=2e^{2}/h$ is the quantum conductance (2 accounts for the spin
degeneracy). The zero-temperature conductance can be rewritten as $G_{\xi
}(E_{F},0)=MG_{0}\int_{-\pi /2}^{\pi /2}T_{\xi }(E_{F},\alpha )\cos \alpha
d\alpha $, where $M=(|E_{F}|/E_{0})(W/2\pi l_{0})\equiv M_E M_W$ is half of the number of
the transverse modes and $2MG_{0}$ is the maximal channel conductance per
valley.

\begin{figure}[tb]
\centering
\includegraphics[width=\linewidth]{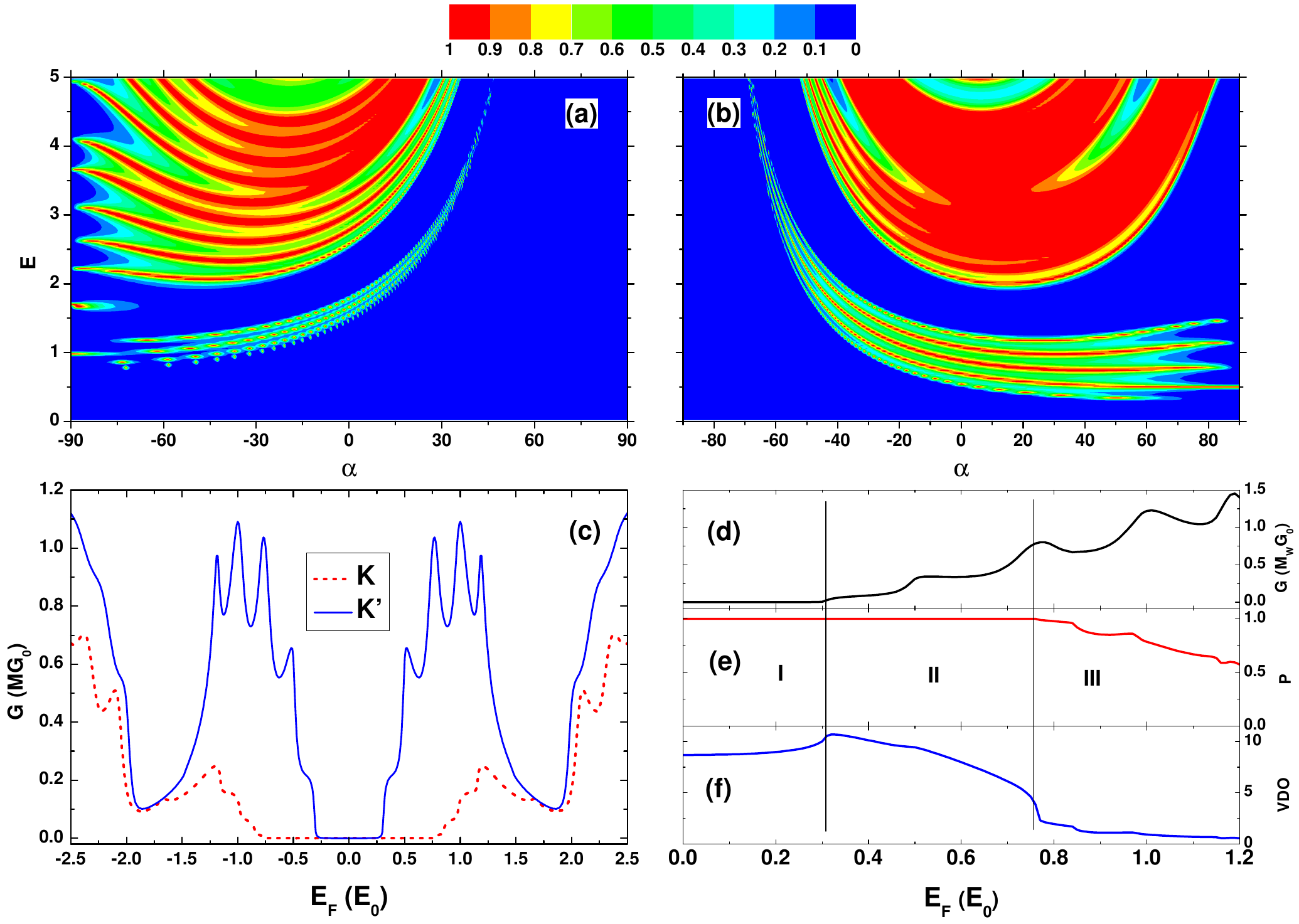}
\caption{(a) and (b) Contour plots of the transmission probability
as a
function of the incident energy and incident angle for valley $K$ and $%
K^{\prime}$. (c)-(f) Valley-resolved conductance, total conductance ($%
G=G_K+G_{K^{\prime}}$), the commonly adopted polarization definition $%
P=(G_{K^{\prime}}-G_K)/G$, and VDO as a function of the
Fermi energy. VDO relates with $P$ by $P=\tanh (VDO\ln10/2)$, but reflects
better the high valley polarization. The parameters are $n=5$, $L_S=L_F=1$, $%
A_S=A_M=1$, $T_P=0$, and $\beta=0$.} \label{fig2}
\end{figure}

In Fig.~\ref{fig2} we present the results for $T_{P}=0$ and $\beta=0$.
As the incident energy decreases, the transmission demonstrates
obvious quasi transparent region and transmission gap [see
Figs.~\ref{fig2}(a) and \ref{fig2}(b)].
The transmission gap is divided into two parts by a resonant region of $(n$-$%
1)$-fold splitting. 
These observations are typical features of a periodic structure,
which come from the $n$-th power terms in Eq.~(\ref{eq1}). As a
result, the valley conductance $G_{K^{\prime}}$ displays obvious
quasi ballistic region, resonant band ($\sim MG_{0}$), and blocked
region ($\sim 10^{-\lambda }MG_{0}$ with $\lambda \geq 4$) in the
Fermi energy windows $|E_{F}|>1.8$, $|E_{F}|\in (0.3,1.8)$, and
$|E_{F}|\in (0,0.3)$, respectively [see Fig.~\ref{fig2}(c)]. For
valley $K$, the transmission is suppressed in a wider energy
window $|E_{F}|\in (0,0.75)$ and the conductance in the resonant
band becomes smaller. This is because the total vector potential
($\mathbf{A}_M+\xi \mathbf{A}_S
$) acting on $K$ electrons is distinct in amplitude from its
counterpart for $K^{\prime }$ electrons.

\begin{figure}[tb]
\centering
\includegraphics[width=\linewidth]{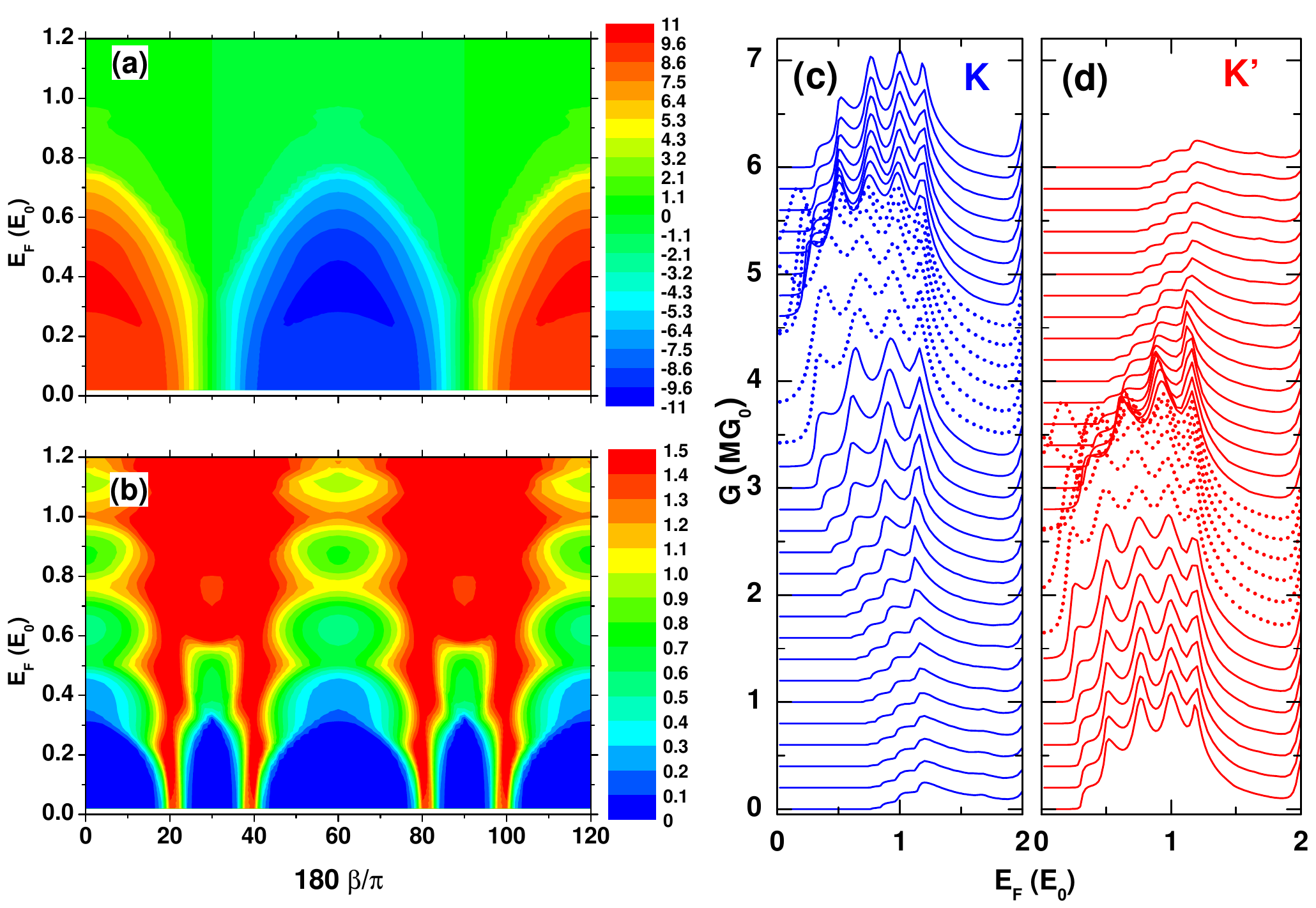}
\caption{(a) and (b) Contour plots of VDO and total conductance as a
function of edge orientation angle and Fermi energy. (c) and (d) $G_K$ and $%
G_{K^{\prime}}$ as a function of $E_F$ at different $\protect\beta$ values.
From bottom to up, $\protect\beta$ increases from 0$^\circ$ to 60$^\circ$
with a step 2$^\circ$. For clarity, the conductance curve at $\protect%
\beta$ is offset by 18$\protect\beta/\protect\pi$. The cases of 34-44$%
^\circ $ (16-26$^\circ$) for valley $K$ ($K^{\prime}$) are highlighted by
dashed curves. Other parameters are the same as in Fig.~2.}
\label{fig3}
\end{figure}

A direct consequence is that, when the Fermi energy falls into the
window II marked in panels (d)-(f)
of Fig.~\ref{fig2}, the current from valley $K$ is almost totally
blocked while the current from valley $K^{\prime}$ possesses the
order of the maximal channel conductance. Such coexistence of the
blocked region of one valley and resonant band of the other
renders the outgoing current a bulk FVPC (contributed mainly by
$K^{\prime}$ electrons). This is clearly reflected from the valley
difference in orders-of-magnitude (VDO) of the outgoing currents,
$VDO=\log(G_{K^{\prime}}/G_K)$ [see Fig.~\ref{fig2}(f)]. In the
window I [$|E_{F}|\in(0,0.3)$], although the valley polarization
is as good as in the window II, the rather small conductance makes
it impossible for either detection or applications.
The generation mechanism is rather similar to that for the nearly
100\% spin-polarized current in half-metals,\cite{half-metal}
i.e., the coexistence of metallic nature and insulating nature for
electrons with opposite spin orientations.
Very recently, a FVPC has also been predicted in ferromagnetic silicene
junctions.\cite{silicene}
Due to the intrinsic gap, 
a much simpler single-junction design is sufficient for the FVPC generation.
When the unit cells'
sizes differ, the $\alpha$ and $E$ ranges for resonant tunneling
become smaller meanwhile the transmission is strongly
suppressed.\cite{disorder} This leads to smaller conductance in
narrower resonant bands for both valleys, hence a smaller FVPC in
an operation window at higher energy.

The dependence of the total conductance and valley polarization on
the edge orientation is shown in Fig.~\ref{fig3}. It is evident
that the total
conductance (polarization) is a periodic function of $\beta $ with a period $%
\pi /3$ ($2\pi /3$). In addition, VDO is antisymmetric (symmetric)
with respect to the angle $\beta =\pi /6$ ($\beta =0$).
These observations result from the
dependence of  $A_{Sy}$ $(=A_{S}\cos 3\beta)$ on the edge
orientation\cite{note} and the symmetry $T_{\xi}(-A_{Sy})=T_{-\xi
}(A_{Sy})$. Thus it is sufficient to consider only the
angle interval $\beta \in \lbrack 0,\pi /6]$. Within this range a larger $%
\beta $ leads to a narrower window of Fermi energy where the VDO
is high [Fig.~\ref{fig3}(a)]. Actually, for $\beta =\pi /6$ (the
zigzag direction), the valley polarization completely disappears
because $G_{-\xi }(\beta )=G_{\xi }(\pi /3-\beta )$
[Figs.~\ref{fig3}(c) and \ref{fig3}(d)].
A remarkable FVPC (high VDO with
substantial $G$) can be obtained under $\beta \approx 0$ and
relatively \emph{high} Fermi energies. It can be also achieved
under $\beta \approx \pi/9$
and relatively \emph{low} Fermi energies.
Under such $\beta$ values, electrons in valley $K^{\prime}$ feel
alternate modulations of total vector potential $\mp
A_S\cos3\beta$ along the $x$ direction, which lead to an almost
transparent transmission [dashed curves in Fig.~\ref{fig3}(d)]. In
the following we will focus only on the armchair edge ($\beta
=0$).

\begin{figure}[tb]
\centering
\includegraphics[width=\linewidth]{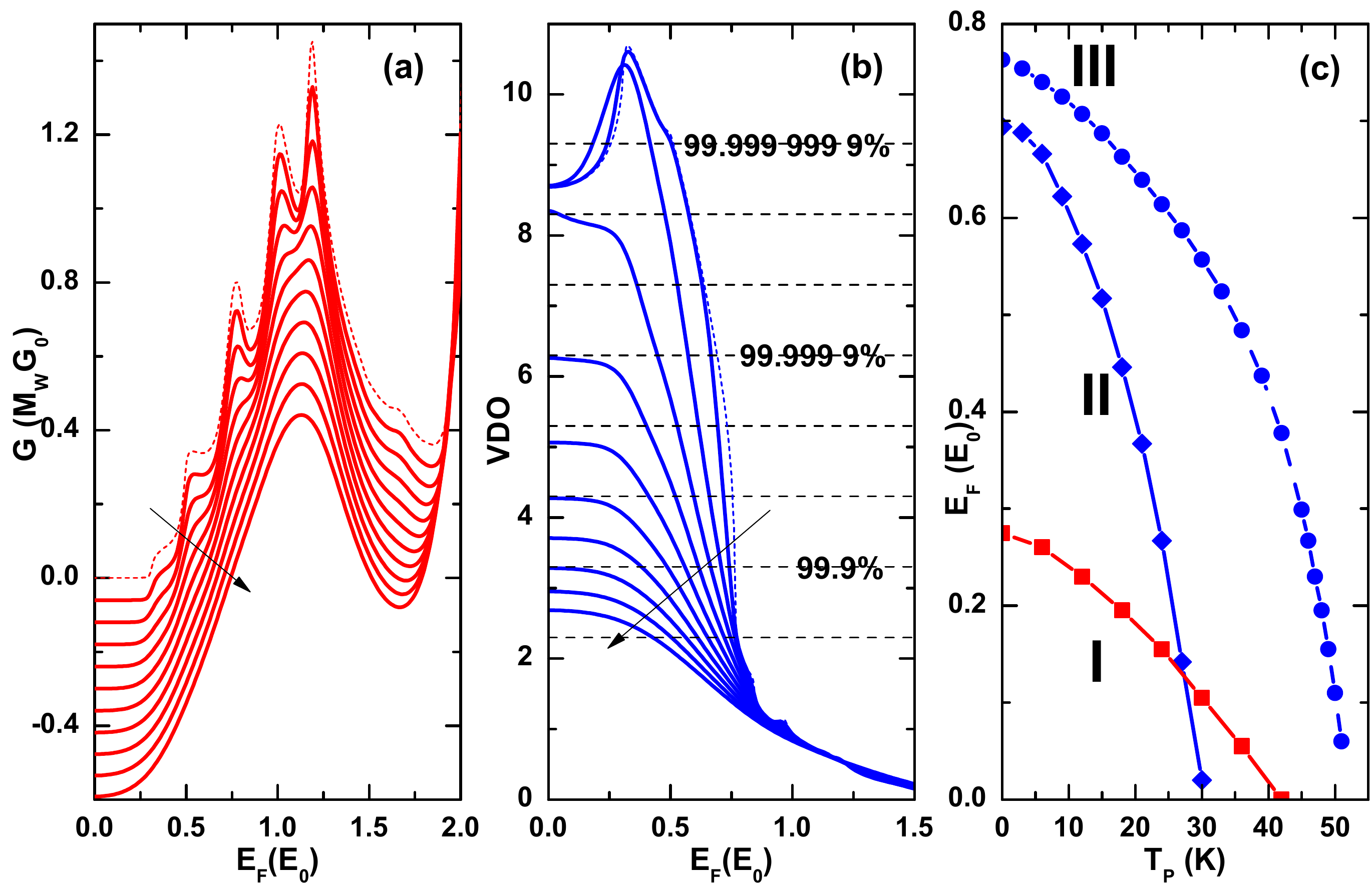}
\caption{(a) Total conductance and (b) VDO as a function of the
Fermi energy at various temperatures. Along the arrow, the
temperature increases from zero to 60$K$ with a step 6$K$. For
clarity, the conductance curves at $T_P$ are offset by
$T_P$/100$K$. The corresponding values of $P$ are marked by dashed
lines. (c) The two Fermi energy
boundaries between the three windows as a function of the
temperature.
The second boundary is shown for $P$ higher than 0.999 999 ($%
\bullet$) or 0.999 999 999 ($\boldsymbol\diamond$).
Other parameters are the same as in Fig.~2. }
\label{fig4}
\end{figure}

We now consider the effect of temperature, which is shown in
Fig.~\ref{fig4}. The rich oscillations in the conductance
spectrum at zero temperature are gradually smeared out as the
temperature increases.
Generally, the valley currents decrease (increase) with the temperature
in the resonant bands (blocked regions).
This leads to increasing total current in window I and decreasing
VDO in window II (see Fig.~\ref{fig4}(a) (b)).
As a result, both windows I and II become smaller and disappear with the increase of
temperature (see Fig.~\ref{fig4}(c)).
These behaviors can be understood from
Eq.~(\ref{eq3}). Valley currents at a finite temperature $T_{P} $
are determined by the zero-temperature valley currents in an
estimated energy range $ \sim(E_{F}-5T_{P},E_{F}+5T_{P})$. In this
range a dramatic variation of $G_{\xi }(T_{P}=0)$ with energy
results in an obvious temperature effect. Nevertheless, we can
still obtain a bulk FVPC in a wide range of Fermi energy at a low
temperature [see Fig.~\ref{fig4}(c)].

\begin{figure}[tb]
\centering
\includegraphics[width=\linewidth]{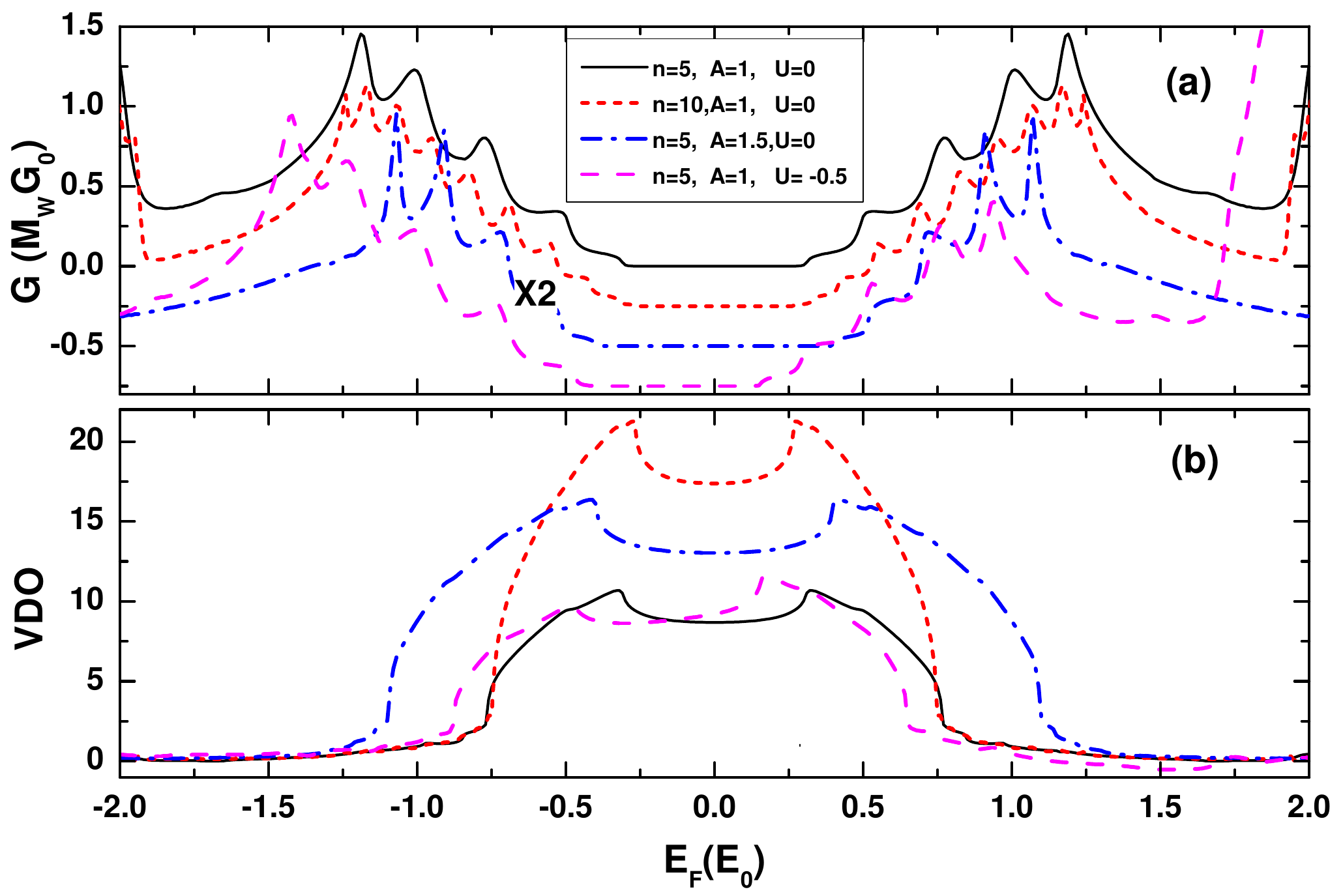}
\caption{(a) Total conductance and (b) VDO as a function of the
Fermi energy for various structural parameters
(where $A_S=A_M=A$) indicated in
the figure.  In (a) the conductance curves are shifted by a step
0.25 and the curve for $A=1.5$ is enlarged by two times for
clearness. $T_P=0$ and $\protect\beta=0$.} \label{fig5}
\end{figure}

The tunability of a FVPC source is
desirable for valleytronic applications. In Fig.~\ref{fig5}, we
examine the influences of structural parameters such as the number
$n$ of superlattice units and the strain strength $A_S$. For
simplicity, the ratio $A_S:A_M$ and $L_S:L_F$ are fixed at $1:1$.
When $n$ increases from $5$ to $10$, the total conductance is
lowered slightly and more conductance peaks appear [see
Eq.~(\ref{eq1})]. As the optimal VDO increases greatly, we can obtain bulk FVPC at
higher temperatures in an almost unchanged energy window. However,
the choice of $n$ should guarantee that the total length is
smaller than the electron mean free path and valley coherent
length. With the increasing of $A_S$, bulk FVPC can appear in a
wider Fermi energy window and the optimal VDO also increases,
while the conductance decreases greatly. Thus the strain (and
magnetization) strength should be chosen as moderate. All the
results above are obtained under electron-hole symmetry. If this
symmetry is broken by a common negative gate voltage on all FM
strips, the conductance in the hole (electron) region increases
(decreases) but the VDO decreases (increases). Both the
conductance and VDO curves shift towards the hole region. As a
result, the inclusion of a negative (positive) electric potential
is helpful for obtaining bulk FVPC for electrons (holes) in lower
energy and at higher temperature.

The injection of a valley-polarized current into a gapped graphene
will lead to a finite Hall voltage.\cite{detect} This mechanism
can be utilized to measure the valley polarization degree of the
filtered current [see Fig.~\ref{fig1}(b)]. Due to the effective
valley dependent gaps\cite{gap} $G_{K^\prime}$ ($G_K$) changes slightly
to $G^\prime_{K^\prime}$ ($G^\prime_K$) in the detection region. The voltage between
the upper (lower) edge and the middle region $V_1$ ($V_2$) is
proportional to $G^\prime_K$ ($G^\prime_{K^\prime}$).\cite{detect} Thus $P$ and VDO
can be respectively reflected by $(V_2-V_1)/(V_2+V_1)$ and
$\log(V_2/V_1)$, especially when the gap is small. A rather huge
positive (negative) $\log(V_2/V_1)$ is an evidence for that we
obtain a FVPC of valley $K^\prime$ ($K$).

In summary, we have demonstrated
that a strained graphene under the modulation of a finite magnetic
superlattice can generate a bulk FVPC. The FVPC appears in a wide
range of edge orientation angles and temperature. The underlying
mechanism is the coexistence of the metallic resonant band and
insulating transmission-blocked region of two valleys in a certain Fermi
energy window. Such a mechanism implies that superlattices
consisting of any valley filtering structures can be used to
generate a FVPC. The FVPC can be effectively controlled by tuning
the structural parameters, and may be used as a high-quality
current source of bulk valleytronics. A valley-resolved Hall
configuration is suggested to measure the valley polarization
degree of the filtered current.

F.Z. acknowledges support from the NSFC Grant No. 11174252. Y.G.
acknowledges support from the NSFC Grant No. 11174168 and the 973
Program Grant No. 2011CB606405.




\begin{references}
\bibitem{wallace1947band}
P. Wallace, Phys. Rev. \textbf{71}, 622 (1947).

\bibitem{semenoff1984condensed}
G.W. Semenoff, Phys. Rev. Lett. \textbf{53}, 2449 (1984).

\bibitem{valve}
A. Rycerz, J. Tworzyd, and C. W. J. Beenakker,
Nature Phys. \textbf{3}, 172 (2007).

\bibitem{detect}
D. Xiao, W. Yao, and Q. Niu,
Phys. Rev. Lett. \textbf{99}, 236809 (2007).

\bibitem{gap}
F. Zhai and K. Chang,
Phys. Rev. B \textbf{85}, 155415 (2012);
D. Moldovan, M.R. Masir, L. Covaci, and F.M. Peeters,
\emph{ibid.} \textbf{86}, 115431 (2012).








\bibitem{warping1}
J.M. Pereira Jr., F.M. Peeters, R.N. Costa Filho, and G.A. Farias,
J. Phys.: Condens. Matter \textbf{21}, 045301 (2009).

\bibitem{strain}
F. Zhai, X.F. Zhao, K. Chang, and H.Q. Xu,
Phys. Rev. B \textbf{82}, 115442 (2010);
A. Chaves, L. Covaci, Kh.Yu. Rakhimov, G.A. Farias, and F.M. Peeters,
\emph{ibid.} \textbf{82}, 205430 (2010);
T. Fujita, M.B.A. Jalil, and S.G. Tan,
Appl. Phys. Lett. \textbf{97}, 043508 (2010);
F. Zhai and L. Yang, \emph{ibid.} \textbf{98}, 062101 (2011).

\bibitem{linedefect}
D. Gunlycke
and C.T. White, Phys. Rev. Lett. \textbf{106}, 136806 (2011); L.
Jiang, G. Yu, W. Gao, Z. Liu, and Y. Zheng, Phys. Rev. B
\textbf{86}, 165433 (2012); J.N.B. Rodrigues, N.M.R. Peres, and
J.M.B. Lopes dos Santos, \emph{ibid.} \textbf{86}, 214206 (2012).

\bibitem{RMP} C. W. J. Beenakker, Rev. Mod. Phys. \textbf{80}, 1337
(2008).

\bibitem{RMP1} A. H. Castro Neto, F. Guinea, N. M. R. Peres, K. S.
Novoselov, and A. K. Geim, Rev. Mod. Phys. \textbf{81}, 109
(2009).


\bibitem{pseudo}
V.M. Pereira and A.H.C. Neto, Phys. Rev. Lett.
\textbf{103}, 046801 (2009).

\bibitem{WL}
J. Tworzyd{\l}o, B. Trauzettel, M. Titov, A. Rycerz, and C.W.J.
Beenakker,
Phys. Rev. Lett. \textbf{96}, 246802 (2006).

\bibitem{tmm}
M. Born and E. Wolf, \emph{Principles of Optics: Electromagnetic
Theory of Propagation, Interference and Diffraction of Light}
(Pergamon, Oxford, 1964).


\bibitem{previous}
In Phys. Rev. Lett.~\textbf{80},
2677 (1998), Pereyra derived a similar formula where $a_n$ and
$b_n$ are Chebyshev polynomials of the second kind in variables
$a$, $b$, $c$, and $d$. That formula assumes $d=a^*$ and $c=b^*$
and holds only for the case $k_S^2>0$.

\bibitem{phonon}
S.V. Morozov, K.S. Novoselov, M.I. Katsnelson, F. Schedin, D.C. Elias, J.A. Jaszczak, and A.K. Geim,
Phys. Rev. Lett. \textbf{100}, 016602 (2008);
J.H. Chen, C. Jang, S. Xiao, M. Ishigami, and M.S. Fuhrer,
Nature Nanotechnol. \textbf{3}, 206 (2008).

\bibitem{LB}
M. B{\"u}ttiker, Y. Imry, R. Landauer, and S. Pinhas,
Phys. Rev. B \textbf{31}, 6207 (1985).

\bibitem{half-metal}
R.A. de Groot, F.M. Mueller, P.G. van Engen, and K.H.J. Buschow,
Phys. Rev. Lett. \textbf{50}, 2024 (1983);
J.H. Park, E. Vescovo, H.J. Kim, C. Kwon, R. Ramesh, and T. Venkatesan,
Nature \textbf{392}, 794 (1998).

\bibitem{silicene}
T. Yokoyama, Phys. Rev. B \textbf{87}, 241409(R) (2013).

\bibitem{disorder}
N. Abedpour, A. Esmailpour, R. Asgari, and M.R.R. Tabar,
Phys. Rev. B \textbf{79}, 165412 (2009).

\bibitem{note}
Note that the change is not due to the edge's
microscopic termination which alter with the edge orientation,
because the sample under consideration is sufficiently wide hence
bulk states dominate the valley-resolved transport.
Due to this factor and large Fermi wavelength at small $E_F$ (see Ref.~%
\onlinecite{valve}), edge imperfections can be also ignored.


\end{references}
\end{document}